\documentclass{optica-article}

\journal{opticajournal} 

\articletype{Research Article}

\usepackage{float}

\begin{document}

\title{Demonstration of an Integrated Terahertz Band-Stop Filter Using an Apodized Bragg Grating}

\author{Ali Dehghanian,\authormark{1,3} Walid Gomaa,\authormark{2}  Mohsen Haghighat\authormark{1,3}  Thomas~Darcie\authormark{1}  and Levi~Smith\authormark{1,3,*}}

\address{\authormark{1}Department of Electrical and Computer Engineering, University of Victoria, Victoria, BC, V8P 5C2 Canada\\
\authormark{2}Department of Engineering Physics, Military Technical College, Cairo, Egypt\\
\authormark{3}Centre for Advanced Materials and Related Technology (CAMTEC), University of Victoria, 3800 Finnerty Rd, Victoria, BC, V8P 5C2, Canada.}

\email{\authormark{*}levismith@uvic.ca} 

\begin{abstract*} 
This paper presents the demonstration of an on-chip integrated Terahertz (THz) Apodized Bragg grating (TABG) which functions as band-stop filter with a center frequency of 0.8 THz and a bandwidth of 200 GHz. For experimentation, we integrate the TABG into our THz System-on-Chip to enable wideband (DC - 1.5 THz) device characterization. Using this methodology, we measure the signal transmission through the TABG and find the experimental results align with simulation and theory provides a rejection of approximately 20 dB across the stop-band.
\end{abstract*}

\section{Introduction}

The terahertz (THz) gap spanning from 0.1 to 10 THz has garnered considerable attention because of its unique capabilities across various fields such as communication, sensing, imaging, and spectroscopy \cite{song2011present,siegel2004terahertz,pickwell2006biomedical,chen2019survey,zhu20233,beard2002terahertz}. However, the progress in developing terahertz devices and systems has been challenging due to the absence of integrated system components such as transmitters, communication channels, receivers, and other passive devices such as filters. Band-stop filters play a crucial role in many applications because they enable the suppression of undesired frequencies within a specific spectral range. At frequencies above \mbox{500 GHz}, there are limited experimental results for planar guided-wave band-stop filters \cite{2023RodillaTHzCapCoupled}, but there are a few key examples: in \cite{2023RodillaTHzCapCoupled}, the authors demonstrate a band-stop filter with a center frequency of 0.6 THz and a -3dB bandwidth of $\approx$160 GHz using a single conductor Goubau line loaded with capacitivly-coupled $\lambda/2$ resonators. This filter behaves well; however, Goubau-lines can be limiting when developing an integrated system because the single conductor is incompatible with two-terminal elements without necessitating a mode converter. Next, others have developed THz band-stop filters using a dielectric Bragg grating with a center frequency of 0.31 THz and a -3dB bandwidth of $\approx$20 GHz\cite{Gao2021_THzfilt}. Again, their filter works well, but, as with most non-transverse electromagnetic (TEM) waveguides, the finite single-mode bandwidth can lead to dispersion in a wideband system. Next, in our prior work we loaded a transmisission line with split-ring resonator elements \cite{smith2021characterization}. Using this method we created a band-stop filter which possessed a center frequency of 0.51 THz and a -3dB bandwidth of $\approx$100 GHz. These SRR filters are useful, but the design procedure is not systematic, the filter roll-off rates are sub-optimal, and higher-order resonator modes must be considered \cite{smith2021characterization}. Lastly, we note that there are several non-planar guided-wave Bragg gratings which have been demonstrated. In \cite{2017_Hollow_BG}, a corrugated circular waveguide was used as a Bragg grating which obtained a center frequency was 0.14 THz and the -3dB bandwidth was $\approx$9 GHz.  In \cite{dong_versatile_2022}, a two-wire waveguide was periodically etched to fabricate the grating which achieved a center frequency of 0.53 THz and the -3dB bandwidth was $\approx$1 GHz.

This work expands upon the aforementioned list of THz band-stop filters by performing the experimental demonstration of an integrated Terahertz Apodized Bragg Grating (TABG) proposed in \cite{gomaa2022design}. Previously the TABG was analyzed via simulations, but no experimental results were presented. Also, we compliment the theory by adding periodic filter concepts from microwave engineering. The TABG was inspired by apodized Fiber Bragg Gratings (FBGs) found in optical communication systems which consist of alternating refractive indices (and wave impedance) along the propagation direction. FBGs experience detrimental side-lobes in the reflection spectrum which are proportional the FBG strength (difference of refractive indices). The magnitude of the side-lobes can be reduced by gradually tapering the FBG strength with an apodization profile. The TABG analogously has a periodic modulation of the characteristic impedance and similarly benefits from gradual tapering of the grating strength.


To characterize the TABG we use our integrated THz System-on-chip (TSoC) platform which combines the THz transmitter, device-under-test (i.e., TABG), and THz receiver onto a single wafer. The TSoC platform consists of planar circuitry which is lithographically defined on an ultra-thin (1 $\mu$m) Si$_3$N$_4$ substrate to ensure signals exhibit low loss and low dispersion at THz frequencies. The ultra-thin substrate is a key requirement to perform wideband measurements otherwise radiation loss into the substrate becomes very problematic. We have used the TSoC platform in previous works to investigate several other THz components such as split-ring resonators \cite{smith2021characterization}, low-pass filters \cite{g2020terahertz}, tapers \cite{smith2021tapered}, and power dividers \cite{gomaa2020terahertz}. Overall, we have found the results from TSoC experiments to closely align with simulation and theory, thus we expect our experimental methodology to be suitable for investigating the TABG.

\begin{figure}[h]
  \centering
  \includegraphics[width=0.8\textwidth]{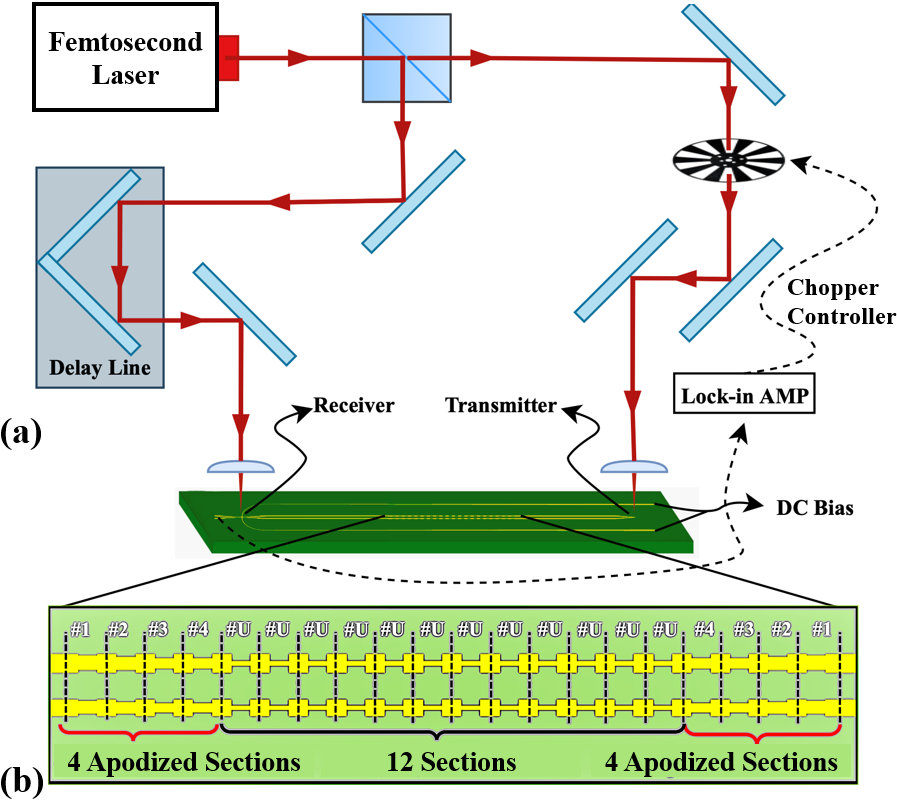}
  \caption{The experimental setup and TABG. (a) The optical path of the transmitter is modulated by the optical chopper, while the receiver beam passes through an optical delay line. The transmitter has a DC bias applied (24V), and the receiver is connected to a lock-in amplifier. The bias lines for the transmitter and receiver are separated by a gap that blocks DC signals. (b) illustrates the TABG filter and its distinct sections.}
  \label{fig:exp}
\end{figure}

\section{Design}
\begin{figure}[H]
  \centering
  \includegraphics[width=5.5in]{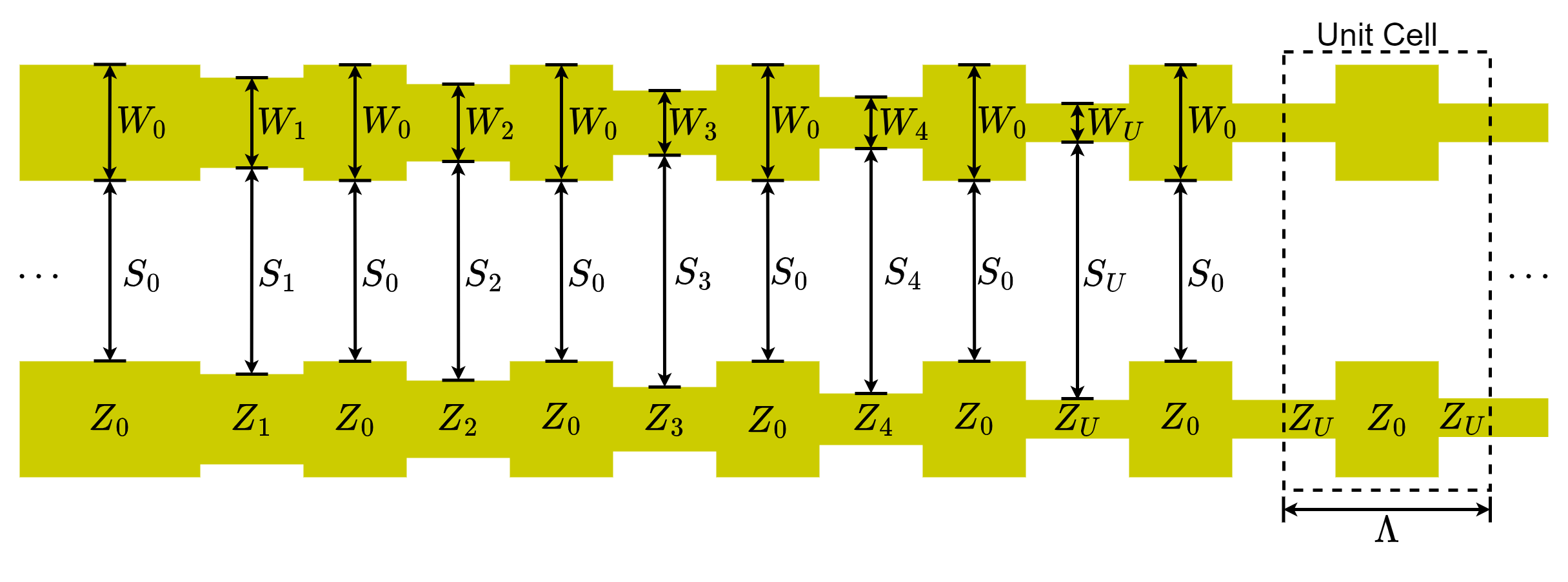}
  \caption{Apodization sections and grating unit cell where $\Lambda = 165 \mu m$. Dimensions and characteristic impedances are found in Table \ref{tab:TABG_dimensions}.}
  \label{fig:apod_sections}
\end{figure}

In this paper, we perform the experimental verification of the TABG shown in Fig. \ref{fig:exp} \cite{gomaa2022design}. Specifically, we focus on a TABG where the stop-band is centered at $f_c$ = 0.8 THz. The filter consists of N = 12 unit cell sections and 4 apodization sections which provide a gradual change in geometry and characteristic impedance between the feedlines and the grating unit cells. \mbox{Figure \ref{fig:apod_sections}} illustrates an annotated portion of the TABG. The TABG response is defined by the reflections caused by characteristic impedance discontinuities along the grating. Thus knowledge of the characteristic impedance of each section is necessary to predict the filter response. In this work, the characteristic impedance of each section is obtained by performing full-wave simulations using ANSYS HFSS at 0.8 THz to ensure accurate values which include the impact of the thin Si$_3$N$_4$ substrate. Table \ref{tab:TABG_dimensions} tabulates the dimensions and simulated characteristic impedances of each section. Next, the grating period, $\Lambda$, required to obtain a specified center frequency, $f_c$,is calculated using:
\begin{equation}
    \Lambda = \frac{c}{2 f_c \sqrt{\varepsilon_{re}}},
    \label{eqn:grating_period}
\end{equation}
where $c$ is the speed of light, $\varepsilon_{re}$ is the effective relative permittivity of the propagating mode. Again, from simulation (ANSYS HFSS), we have found $\varepsilon_{re} \approx 1.3$ for the TABG CPS configuration, then using (\ref{eqn:grating_period}), we find \mbox{$\Lambda$ = 165 $\mu$m}.  Next, the fractional bandwidth, $\Delta f/f_c$, is dependent on the characteristic impedance of adjacent cells and is calculated by \cite{Orfanidis_2011}:
\begin{equation}
    \frac{\Delta f}{f_c} = \frac{4}{\pi} \sin^{-1} \left( \frac{Z_{n+1}-Z_n}{Z_{n+1}+Z_n} \right),
    \label{eqn:frac_bandwidth}
\end{equation}
using (\ref{eqn:frac_bandwidth}), with $Z_0$ and $Z_U$ (Table \ref{tab:TABG_dimensions}), we find \mbox{$\Delta f = 0.18$ THz}.

\begin{table}[H]
    \centering
    \caption{TABG dimensions and characteristic impedances for Fig. \ref{fig:apod_sections}}
    \begin{tabular}{|c|c|c|c|c|c|c|} \hline
        $n$ &  0 &  1 & 2& 3 & 4 & U \\ \hline
        W$_n$ [$\mu$m] &  45 & 35 & 30 & 25 & 20 & 15 \\ \hline
        S$_n$ [$\mu$m]&  70 &  80 & 85& 90 & 95 & 100 \\ \hline
        Z$_n$ [$\Omega$]&  234 &  260 & 274& 290 & 311 & 332 \\ \hline
    \end{tabular}
    \label{tab:TABG_dimensions}
\end{table}

\section{Theory}
\label{sec:theory}
In microwave engineering, the Bragg filter of \cite{gomaa2022design} is commonly referred to as a periodic filter which can be described by a dispersion diagram and Bloch impedance \cite{pozar_microwave_2011}. To calculate these quantities we construct an ABCD matrix of the unit cell depicted in Fig. \ref{fig:apod_sections} as:
\begin{equation}
\begin{split}
\begin{bmatrix}
A & B \\
C & D 
\end{bmatrix}  =
& \begin{bmatrix}
\text{cos} \beta \Lambda/4 & j Z_U \text{sin} \beta \Lambda/4 \\
j Y_U \text{sin} \beta \Lambda/4 & \text{cos} \beta \Lambda/4 
\end{bmatrix} \\
\cdot & \begin{bmatrix}
\text{cos} \beta \Lambda/2 & j Z_0 \text{sin} \beta \Lambda/2 \\
j Y_0 \text{sin} \beta \Lambda/2 & \text{cos} \beta \Lambda/2 
\end{bmatrix} \\
\cdot & \begin{bmatrix}
\text{cos} \beta \Lambda/4 & j Z_U \text{sin} \beta \Lambda/4 \\
j Y_U \text{sin} \beta \Lambda/4 & \text{cos} \beta \Lambda/4 
\end{bmatrix},
\end{split}
\end{equation}
where $\beta = \omega \sqrt{\varepsilon_{re}}/c$ is the propagation constant.

\begin{figure}[h!]
  \centering
  \includegraphics[width=4in]{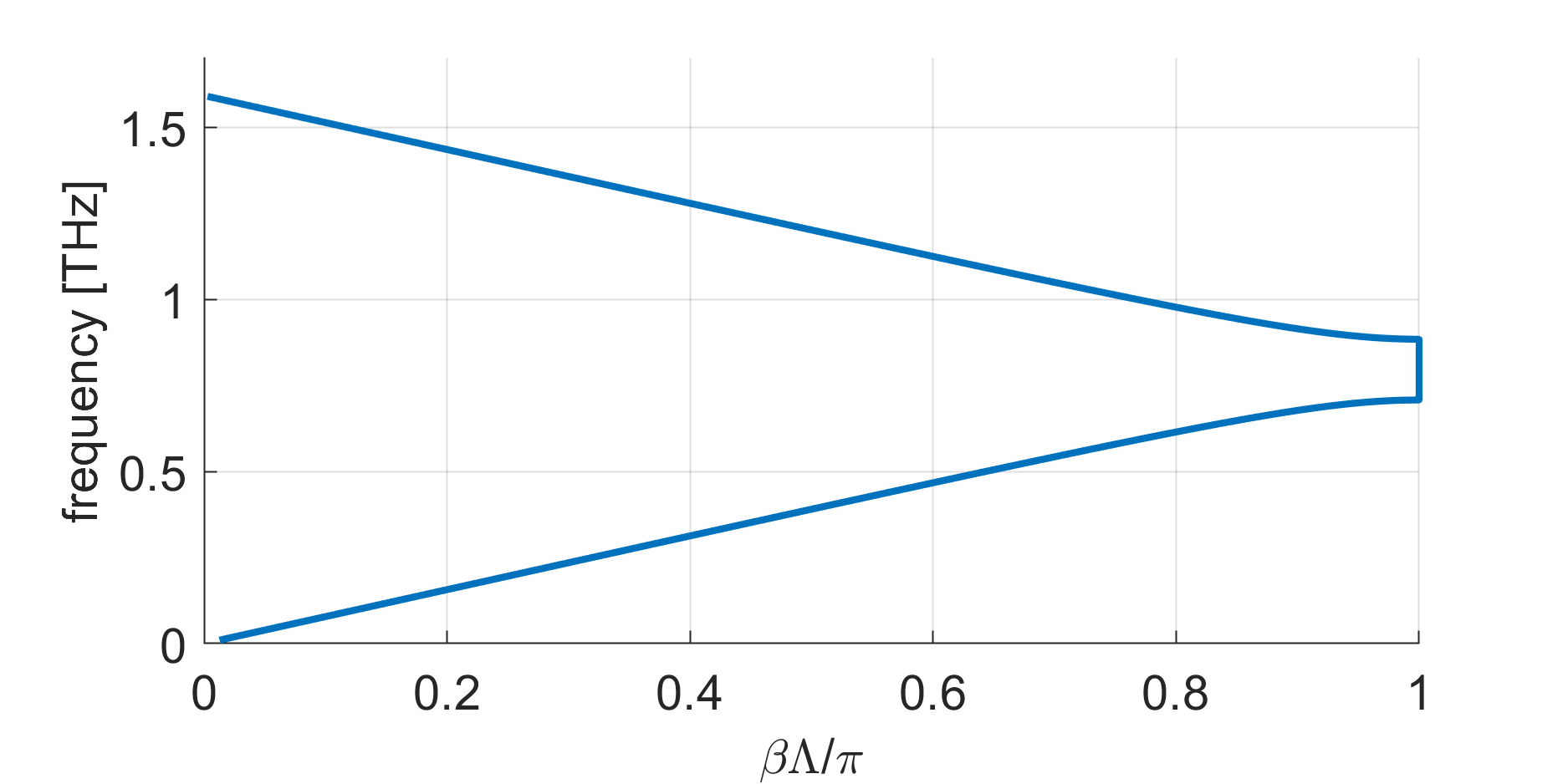}
  \caption{Dispersion diagram for the TABG.}
  \label{fig:dispersion}
\end{figure}

\begin{figure}[h!]
  \centering
  \includegraphics[width=4in]{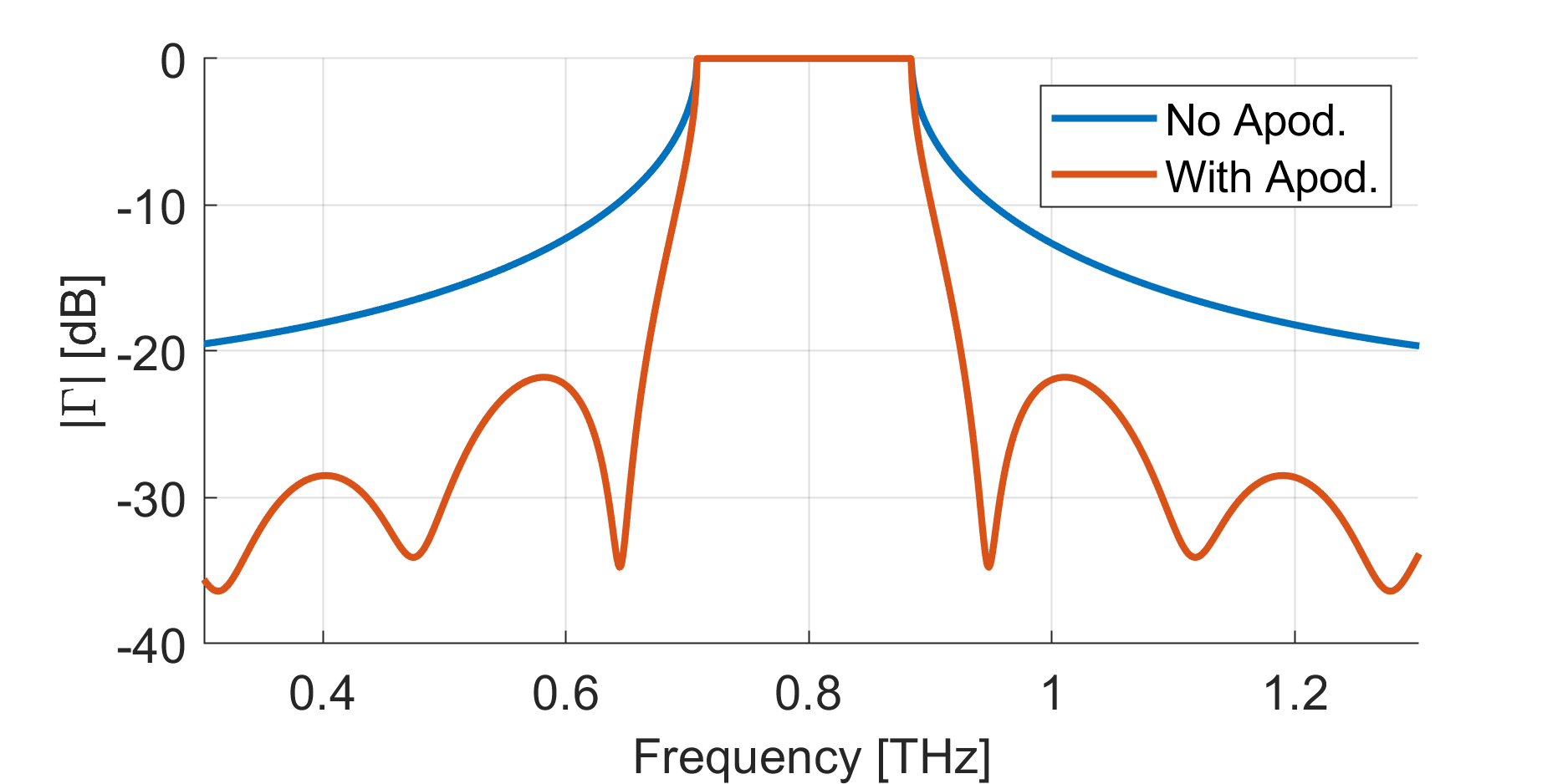}
  \caption{Reflection coefficient at the load of the TABG with and without an apodization section.}
  \label{fig:reflection}
\end{figure}

After constructing the ABCD matrix we can obtain the dispersion diagram and Bloch impedance. The dispersion diagram is calculated using $\beta \Lambda = \text{Imag}\{\mbox{cosh}^{-1}[(A+D)/2]\}$ and the result is plotted in Fig. \ref{fig:dispersion}. This figure illustrates the stopband associated with the infinite periodic structure. We find that the filter has a center frequency at $f_c$ = 0.8 THz and a bandwidth of \mbox{$\Delta f \approx 0.18$ THz} which is in agreement with (\ref{eqn:frac_bandwidth}). Next, we calculate the Bloch impedance using $Z_B = B/\sqrt{A^2 - 1}$ then calculate the reflection coefficient between an infinite periodic filter and a load impedance from:
\begin{equation}
    \Gamma = \frac{Z_L-Z_B}{Z_L+Z_B}.
    \label{eqn:ref}
\end{equation}
We use (\ref{eqn:ref}) to illustrate the impact of apodization. If we negate the apodization sections, then $Z_L = Z_0$ which will result in a mismatch causing a less desirable reflection spectrum. Alternatively, if we add an apodization section then we can significantly improve the matching and reduce the reflections. We model apodization section by cascading 4 ABCD matrices to represent each apodization cell in \mbox{Fig. \ref{fig:apod_sections}}. The result of this procedure is plotted in \mbox{Fig. \ref{fig:reflection}} where we see the introduction of apodization sections improves the reflection response of the filter and we expect to see a sharp roll-off at the band-edge of the transmitted signal in our experimental results.

\section{Simulation}
\begin{figure}[H]
  \centering
  \includegraphics[width=0.99\textwidth]{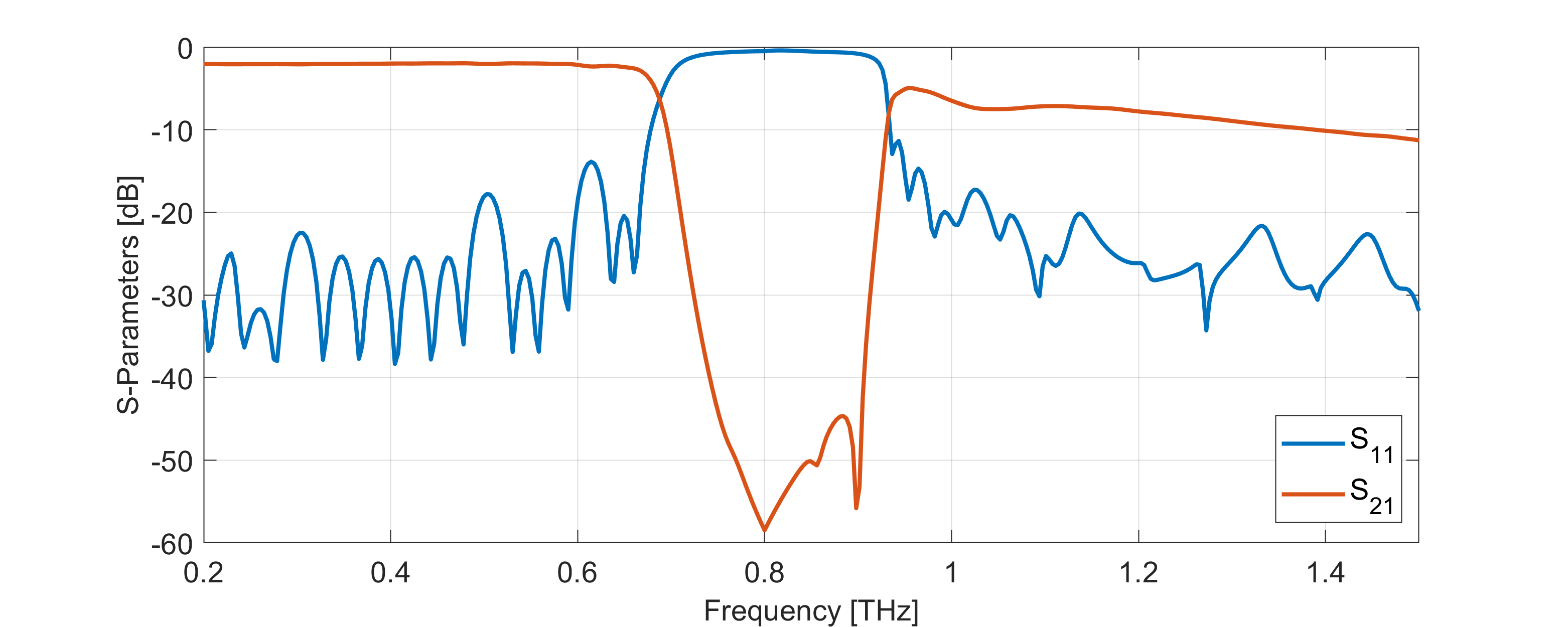}
  \caption{Simulated S-Parameters. $f_c$ = 0.815 THz and the -3 dB bandwidth is 0.22 THz.}
  \label{fig:sim_sparams}
\end{figure}

A full-wave frequency domain simulation was performed using ANSYS HFSS to characterize the scattering parameters of the TABG (Fig. \ref{fig:sim_sparams}). In the simulation, the material parameters are given by $\varepsilon_r$ = 7.6, $\sigma_{Si_3N_4}=0$, $\mu_r$ = 1, \mbox{tan $\delta_e$ = 0.00526} for the Si$_3$N$_4$ substrate\cite{Cataldo_Silicon_nitride_properties_2012} and $\sigma_{Au}=4.1 \times 10^7$ S/m for the gold conductors. All geometric parameters are the same as the fabricated device (see Fig. \ref{fig:apod_sections} and Table \ref{tab:TABG_dimensions}). The simulated center frequency was found to be $f_c$ = 0.815 THz and the -3 dB bandwidth was found to be 0.22 THz which illustrates reasonable agreement between theory and simulation. We do not expect perfect agreement between Fig. \ref{fig:reflection} ($| \Gamma |$) and Fig. \ref{fig:sim_sparams} ($|S_{11}|$) because the reflection coefficient of (\ref{eqn:ref}) calculates the reflection between an infinite periodic structure and a constant load impedance, whereas the simulation accounts for the finite length grating and the frequency-dependent characteristic impedance. The insertion loss below the Bragg regime is less than 2 dB, whereas above the Bragg regime, the insertion loss increases from 5 dB at 0.95 THz up-to 11 dB at 1.5 THz. The increased insertion loss originates from diffractive grating radiation \cite{cheben_subwavelength_2018}. This effect is discussed and illustrated in Appendix A.

\section{Methods}
To perform the experimental characterization of the TABG we use a modified THz Time Domain Spectroscopy (THz-TDS) setup as shown in \mbox{Fig. \ref{fig:exp}(a)}. An optical pulse train is generated by a 780 nm, 20 mW, 80 femtosecond laser and then is divided into two beams that are directed towards the transmitter and receiver photoconductive switches (PCSs) which are thin films of LT-GaAs measuring \mbox{70 \textmu m × 40 \textmu m × 1.8 \textmu m}. The procedure to fabricate a grid of PCS's is described in \cite{Rios2015_bowtie_PCA, g2020terahertz}. For the transmitter, the optical beam passes through a optical chopper before being focused onto the PCS. The receiver beam passes through a mechanical delay line before getting focused onto the PCS. The transmitter has a bias voltage applied (24V) and the receiver is connected to a lock-in amplifier. Translation of the mechanical delay line and measurement of the lock-in amplifier current reconstructs a signal proportional to the transmission characteristics of the TABG.

\begin{figure}
  \centering
  \includegraphics[width=0.98\textwidth]{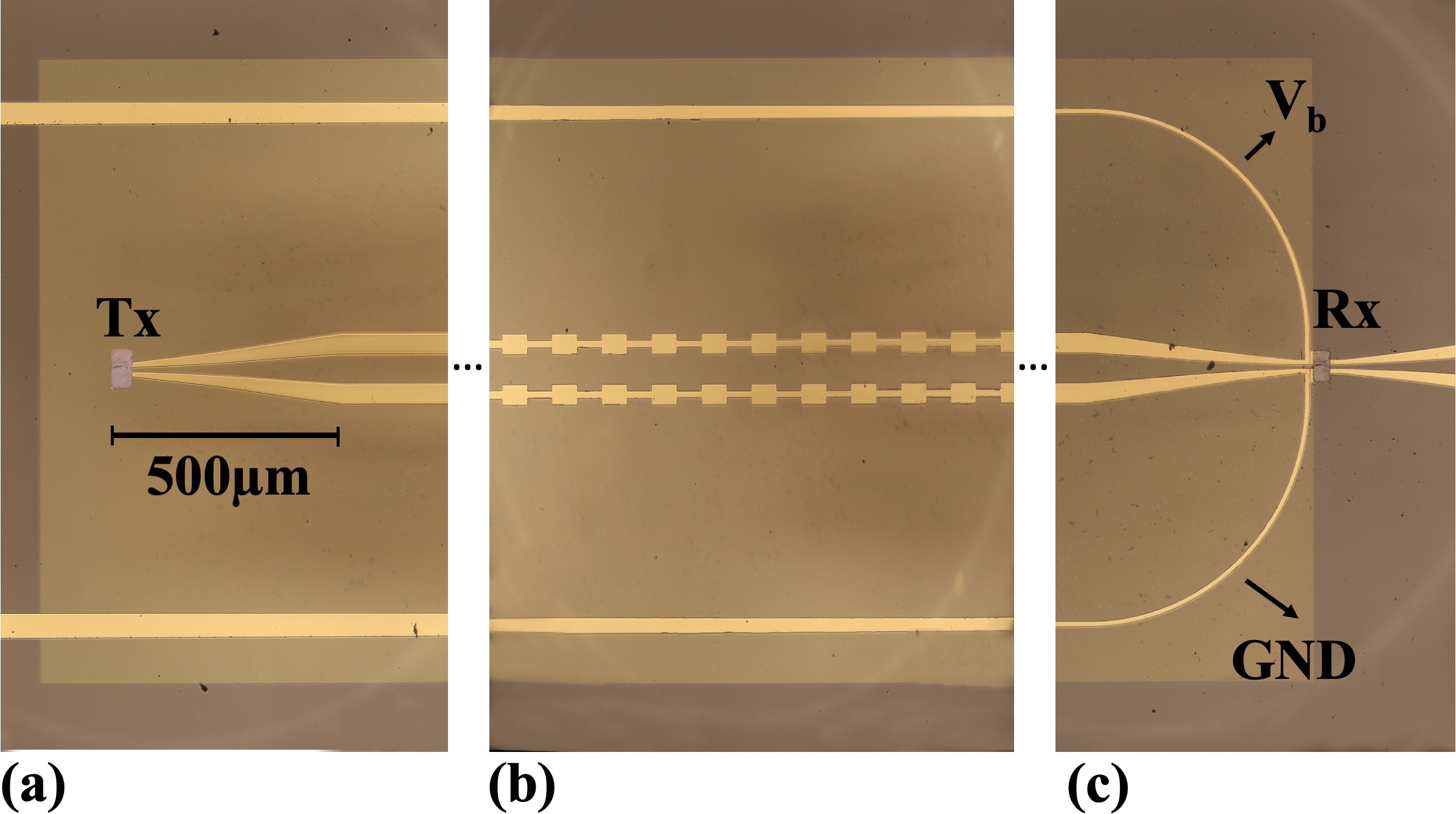}
  \caption{TSoC with TABG. (a) illustrates a LT-GaAs PCS (Tx) on a gold transmission line on a 1\textmu m-thin Si$_{3}$N$_{4}$ membrane. In (b), the TABG is displayed. (c) shows the LT-GaAs PCS receiver (Rx), coupling section, and the DC bias lines.}
  \label{fig:fab_TSoC}
\end{figure}

Figure \ref{fig:fab_TSoC}(a) illustrates the transmitter which is placed on top of a lithographically-defined 200 nm gold CPS TL situated on a 1 \textmu m Si$_3$N$_4$ thin layer. We use a thin Si$_{3}$N$_{4}$ membrane as the substrate to enable the transmission of sub-picosecond pulses over centimeter-scale distances. If a thicker substrate were used, radiation from the substrate would cause significant losses and dispersion \cite{smith2019demonstration}. We have successfully used this method in several other works \cite{smith2021characterization,g2020terahertz,smith2021tapered,gomaa2020terahertz}. After the transmitter PCS, we taper the CPS TL to achieve a wider cross-section (W = 45 µm and S = 70 µm) which has lower attenuation \cite{smith2021tapered}. In Fig. \ref{fig:fab_TSoC}(b), the TABG structure is displayed under a microscope. Fig. \ref{fig:fab_TSoC}(c) portrays the receiver section, which comprises of a DC block, DC bias lines for the Tx PCS, and bias lines to connect the Rx PCS to the lock-in amplifier.

\section{Results and Discussion}

\begin{figure}
  \centering
  \includegraphics[width=0.98\textwidth]{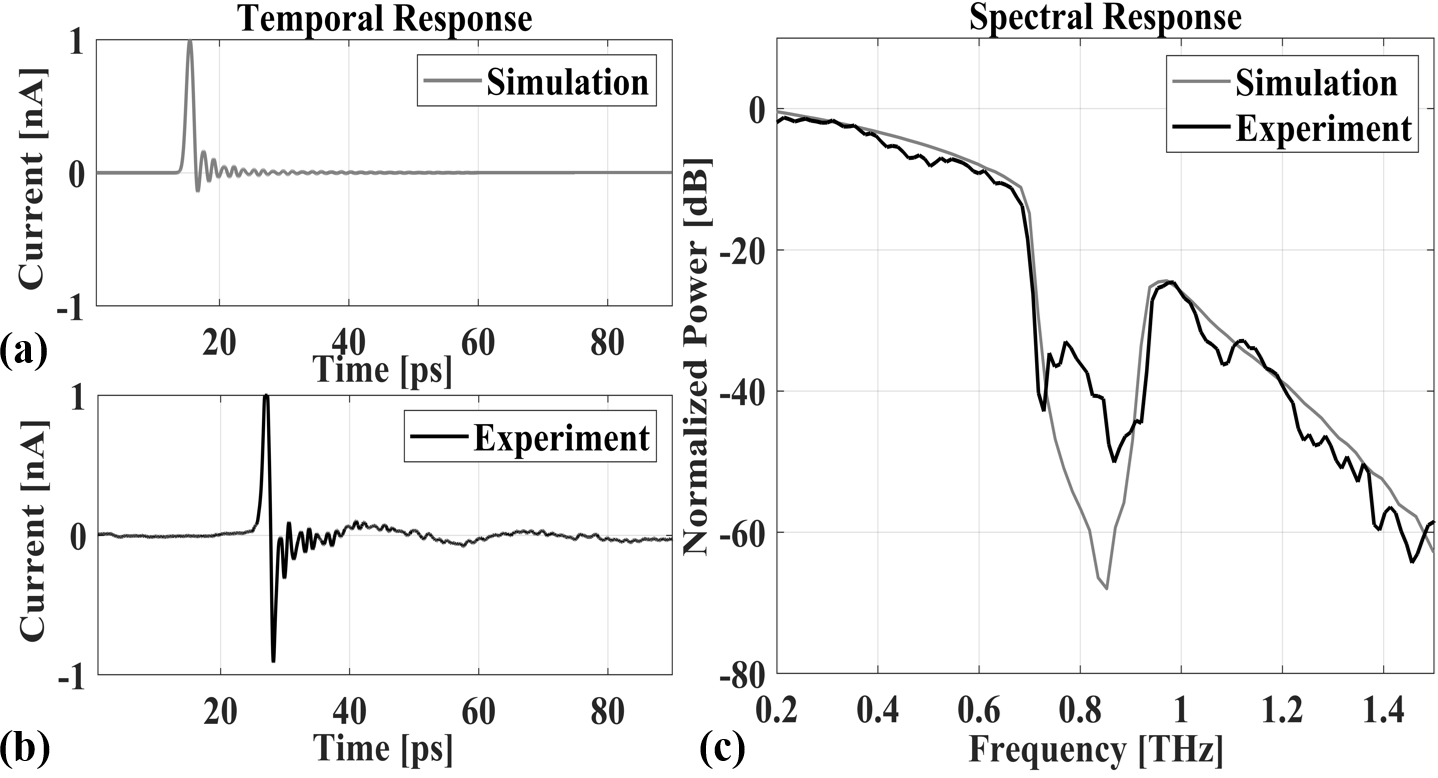}
  \caption{The experimental result of the received THz-bandwidth pulse after propagation through the TABG. (a) The simulated temporal response (ANSYS HFSS). (b) The experimental temporal response. (c) The DFT of the experimental and simulated temporal responses.}
  \label{fig:results}
\end{figure}

Figure \ref{fig:results}(a-b) plot the simulated and experimental temporal response of the received THz-bandwidth pulse after propagating through the TABG. Figure \ref{fig:results}(c) plots the spectral response obtained by applying the Discrete Fourier Transform (DFT) to the temporal response. First, we note that the spectral roll-off associated with both the simulation and experiment is an expected consequence of a finite duration time-domain pulse. Next, we observe good agreement between experiment and simulation which is observed by comparing the stop-band center frequency and bandwidth. We do notice a discrepancy in the stop-band rejection. For the experiment the rejection is $\approx$30 dB, whereas for simulation it is $\approx$45 dB. The difference originates from a few different mechanisms. First, our photolithographic fabrication process is limited to a feature size of $\approx$2 $\mu$m, thus there will not be perfect agreement between the simulated dimensions and the fabricated dimensions. Next, the material parameters will differ between simulation and experiment. While we expect the material parameters to be close, they will not match the fabricated device and will result in differences. Also, it is possible that another unexpected mode coupled across the filter experiencing less attenuation. Lastly, we explain a discrepancy temporal response. The simulation does not use a DC blocking capacitor, thus it contains a DC component. The experiment requires a DC block, therefore we observe the derivative behaviour (high-pass filtering). Note that the cut-off frequency is below our frequencies of interest (0.2 THz) and it is not observed in the spectral response.

\section{Conclusion}
This work presented the experimental validation of a Terahertz Apodized Bragg Grating (TABG) \cite{gomaa2022design}. The TABG was designed to have a center frequency of 0.8 THz and a bandwidth of 0.2 THz. These parameters were confirmed by experiment. Apodization serves to reduce the side-lobes and increase the roll-off rate of the filter which was observed in simulation and experiment. The results showed that the filter had a minimum rejection of approximately 20 dB over the stop-band.

Novelty points: (1) The first demonstration of a fully integrated (transmitter, feedlines, TABG, receiver) on-chip Apodized Bragg filter in the THz gap. (2) The first demonstrated the highest Bragg frequency achieved within the THz gap using a quasi-TEM feedlines. (3) The first demonstration of a Bragg grating constructed from guided-wave TLs on an ultra-thin Si$_3$N$_4$ membrane. (4) We verify that the theory for periodic filter is applicable the grating design. (5) We illustrate the radiative loss mechanism for frequencies above the Bragg bandgap for a CPS grating (Appendix A).

\section*{Funding}
We acknowledge the support of the Natural Sciences and Engineering Research Council of Canada (NSERC).

\section*{Acknowledgments}
This work made use of the 4D LABS core facility at Simon Fraser University (SFU) supported by the Canada Foundation for Innovation (CFI), British Columbia Knowledge Development Fund (BCKDF), and Pacific Economic Development Canada (PacifiCan). We would like to acknowledge CMC Microsystems for the provision of products and services that facilitated this research.

\section*{Disclosures}
The authors declare no conflicts of interest.

\section*{Data availability}
Data underlying the results presented in this paper are not publicly available at this time but may be obtained from the authors upon reasonable request.

\bibliography{main}

\begin{thebibliography}{10}
\newcommand{\enquote}[1]{``#1''}

\bibitem{song2011present}
H.-J. Song and T.~Nagatsuma, \enquote{Present and future of terahertz
  communications,} {\protect\JournalTitle{IEEE transactions on terahertz
  science and technology}} \textbf{1}, 256--263 (2011).

\bibitem{siegel2004terahertz}
P.~H. Siegel, \enquote{Terahertz technology in biology and medicine,}
  {\protect\JournalTitle{IEEE transactions on microwave theory and techniques}}
  \textbf{52}, 2438--2447 (2004).

\bibitem{pickwell2006biomedical}
E.~Pickwell and V.~Wallace, \enquote{Biomedical applications of terahertz
  technology,} {\protect\JournalTitle{Journal of Physics D: Applied Physics}}
  \textbf{39}, R301 (2006).

\bibitem{chen2019survey}
Z.~Chen, X.~Ma, B.~Zhang, Y.~Zhang, Z.~Niu, N.~Kuang, W.~Chen, L.~Li, and
  S.~Li, \enquote{A survey on terahertz communications,}
  {\protect\JournalTitle{China Communications}} \textbf{16}, 1--35 (2019).

\bibitem{zhu20233}
L.~Zhu, S.-H. Shin, R.~Payapulli, T.~Machii, M.~Motoyoshi, N.~Suematsu, N.~M.
  Ridler, and S.~Lucyszyn, \enquote{3-d printed rectangular waveguide 123--129
  ghz packaging for commercial cmos rfics,} {\protect\JournalTitle{IEEE
  Microwave and Wireless Technology Letters}}  (2023).

\bibitem{beard2002terahertz}
M.~C. Beard, G.~M. Turner, and C.~A. Schmuttenmaer, \enquote{Terahertz
  spectroscopy,}  (2002).

\bibitem{2023RodillaTHzCapCoupled}
J.~Cabello-Sánchez, V.~Drakinskiy, J.~Stake, and H.~Rodilla,
  \enquote{Capacitively-coupled resonators for terahertz planar-goubau-line
  filters,} {\protect\JournalTitle{IEEE Transactions on Terahertz Science and
  Technology}} \textbf{13}, 58--66 (2023).

\bibitem{Gao2021_THzfilt}
W.~Gao, W.~S.~L. Lee, C.~Fumeaux, and W.~Withayachumnankul,
  \enquote{{Effective-medium-clad Bragg grating filters},}
  {\protect\JournalTitle{APL Photonics}} \textbf{6}, 076105 (2021).

\bibitem{smith2021characterization}
L.~Smith, V.~Shiran, W.~Gomaa, and T.~Darcie, \enquote{Characterization of a
  split-ring-resonator-loaded transmission line at terahertz frequencies,}
  {\protect\JournalTitle{Optics Express}} \textbf{29}, 23282--23289 (2021).

\bibitem{2017_Hollow_BG}
T.~Ma, K.~Nallapan, H.~Guerboukha, and M.~Skorobogatiy, \enquote{Analog signal
  processing in the terahertz communication links using waveguide bragg
  gratings: example of dispersion compensation,} {\protect\JournalTitle{Opt.
  Express}} \textbf{25}, 11009--11026 (2017).

\bibitem{dong_versatile_2022}
J.~Dong, A.~Tomasino, G.~Balistreri, P.~You, A.~Vorobiov, E.~Charette,
  B.~Le~Drogoff, M.~Chaker, A.~Yurtsever, S.~Stivala, M.~A. Vincenti,
  C.~De~Angelis, D.~Kip, J.~Azana, and R.~Morandotti, \enquote{Versatile
  metal-wire waveguides for broadband terahertz signal processing and
  multiplexing,} {\protect\JournalTitle{Nature Communications}} \textbf{13},
  741 (2022).

\bibitem{gomaa2022design}
W.~Gomaa and T.~Darcie, \enquote{Design and simulation of terahertz apodized
  bragg grating using coplanar stripline transmission line a 1 $\mu$m-thin
  membrane,} in \emph{Journal of Physics: Conference Series,}  vol. 2304 (IOP
  Publishing, 2022), p. 012015.

\bibitem{g2020terahertz}
W.~Gomaa, L.~Smith, V.~Shiran, and T.~Darcie, \enquote{Terahertz low-pass
  filter based on cascaded resonators formed by cps bending on a thin
  membrane,} {\protect\JournalTitle{Optics Express}} \textbf{28}, 31967--31978
  (2020).

\bibitem{smith2021tapered}
L.~Smith, W.~Gomma, H.~Esmaeilsabzali, and T.~Darcie, \enquote{Tapered
  transmission lines for terahertz systems,} {\protect\JournalTitle{Optics
  Express}} \textbf{29}, 17295--17303 (2021).

\bibitem{gomaa2020terahertz}
W.~Gomaa, R.~L. Smith, H.~Esmaeilsabzali, and T.~E. Darcie, \enquote{Terahertz
  power divider using symmetric cps transmission line on a thin membrane,}
  {\protect\JournalTitle{IEEE Access}} \textbf{8}, 214425--214433 (2020).

\bibitem{Orfanidis_2011}
S.~Orfanidis, \emph{Electromagnetic Waves and Antennas} (Online, 2016).

\bibitem{pozar_microwave_2011}
D.~Pozar, \emph{Microwave {Engineering}} (Wiley, 2011), 4th ed.

\bibitem{Cataldo_Silicon_nitride_properties_2012}
G.~Cataldo, J.~A. Beall, H.-M. Cho, B.~McAndrew, M.~D. Niemack, and E.~J.
  Wollack, \enquote{Infrared dielectric properties of low-stress silicon
  nitride,} {\protect\JournalTitle{Optics Letters}} \textbf{37}, 4200--4202
  (2012).

\bibitem{cheben_subwavelength_2018}
P.~Cheben, R.~Halir, J.~H. Schmid, H.~A. Atwater, and D.~R. Smith,
  \enquote{Subwavelength integrated photonics,} {\protect\JournalTitle{Nature}}
  \textbf{560} (2018).

\bibitem{Rios2015_bowtie_PCA}
R.~D.~V. Rios, S.~Bikorimana, M.~A. Ummy, R.~Dorsinville, and S.-W. Seo,
  \enquote{A bow-tie photoconductive antenna using a low-temperature-grown gaas
  thin-film on a silicon substrate for terahertz wave generation and
  detection,} {\protect\JournalTitle{Journal of Optics}} \textbf{17}, 125802
  (2015).

\bibitem{smith2019demonstration}
R.~Smith and T.~Darcie, \enquote{Demonstration of a low-distortion terahertz
  system-on-chip using a cps waveguide on a thin membrane substrate,}
  {\protect\JournalTitle{Optics express}} \textbf{27}, 13653--13663 (2019).

\end{thebibliography}






\section*{Appendix A - Grating radiation above the stopband}

\begin{figure}[H]
  \centering
  \includegraphics[width=0.9\textwidth]{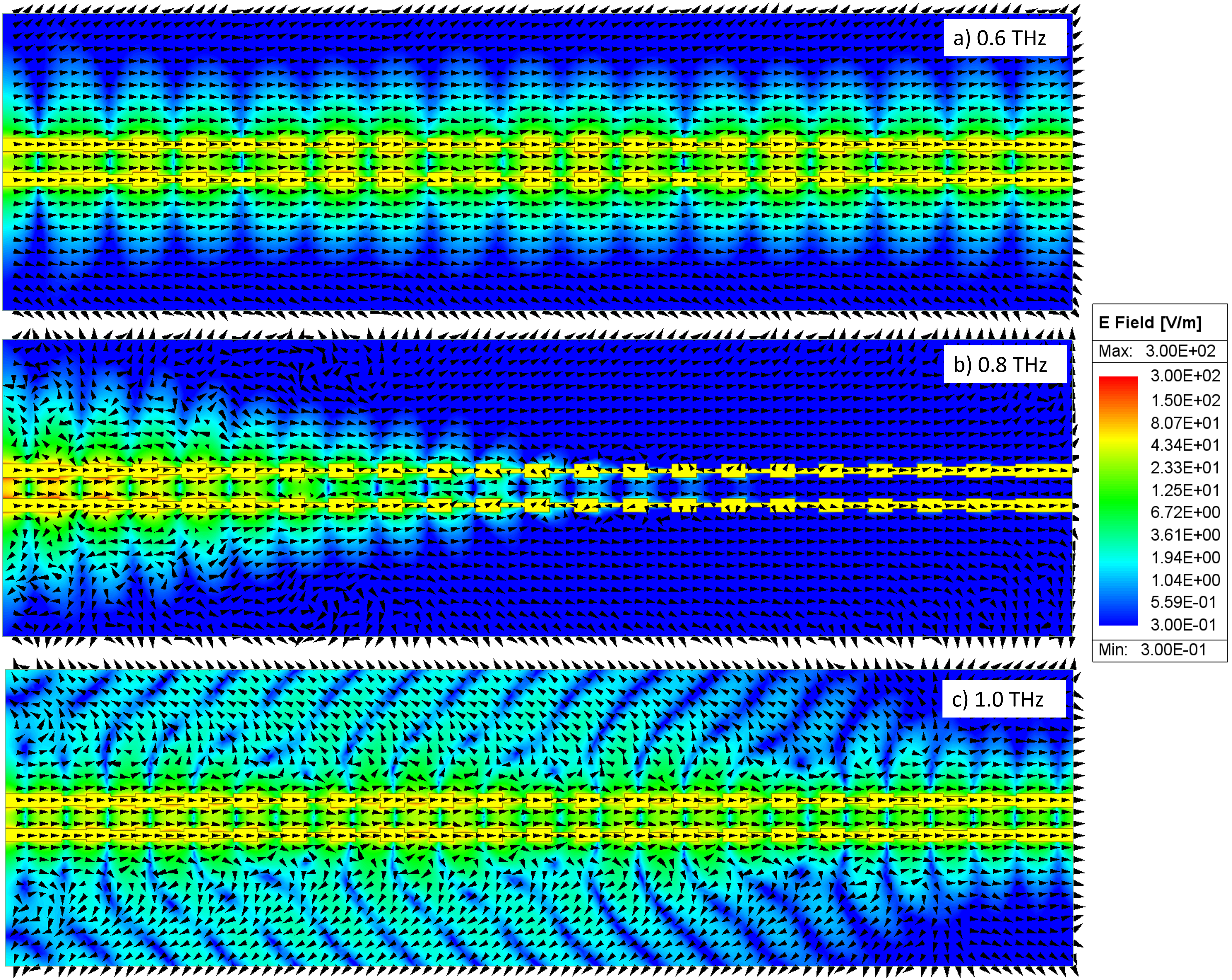}
  \caption{Simulated electric field intensity profiles. Black arrows illustrate the Poynting vector direction. a) Below the stopband, f = 0.6 THz. b) In the stopband, f = 0.8 THz. c) Above the stopband, f = 1.0 THz.}
  \label{fig:sim_efield}
\end{figure}

When operating at frequencies above the Bragg bandgap there is an increased insertion loss which originates from diffractive grating radiation. Figure \ref{fig:sim_efield} illustrates this concept where the black arrows illustrate the direction of the Poynting vector. Note that the color scale is logarithmic and covers three orders of magnitude. Below the bandgap, Fig. \ref{fig:sim_efield}(a), the Poynting vector illustrates a power flow from the left to right with no extra radiative loss. Within the bandgap, Fig. \ref{fig:sim_efield}(b), the incident signal from is reflected such that minimal power is transmitted. Above the bandgap, Fig. \ref{fig:sim_efield}(c), a portion of the reflected wave is back--scattered by grating in the tangential direction which contributes to the additional insertion loss.

\end{document}